\documentstyle[12pt,epsfig]{article} 

\begin{document}

\rightline{UK/99-03}

\vspace{0.3in}

\centerline{\Large \bf Leading soft gluon production}
\centerline{\Large \bf in high energy nuclear collisions
\footnote{Talk presented in the 15th Winter Workshop on Nuclear Dynamics, 
To aprrear in the conference proceeding published by Kluwer Academic 
Publisher. This work is supported in part by the U.S. Department 
of Energy under 
Grant Nos. DE-FG02-93ER40764 and DE-FG02-96ER40989.}}  

\vspace{0.3in}

\centerline{\it Xiaofeng Guo}

\centerline{ \it Department of Physics and Astronomy, 
University of Kentucky}
\centerline{\it Lexington, KY 40506, USA}


\medskip 

\begin{abstract}
The leading soft gluon $p_T$ distribution in heavy ion collisions  
was obtained by Kovner, McLerran, and Weigert after solving classical 
Yang-Mills equations. I show explicitly this result can be understood 
in terms of conventional QCD perturbation theory. I also demonstrate 
that the key logarithm  in their result represents the logarithm 
in DGLAP evolution equations.
\end{abstract}


\section{Introduction}

In ultra-relativistic heavy ion collisions, after initial collisions, the 
parton system will be dominated by gluons \cite{Eskola}. 
Understanding the distribution of soft gluons formed in the initial 
stage of the collision is particularly interesting and important for 
studying the formation of quark-gluon plasma. In terms of conventional 
QCD perturbation
theory, a calculable cross
section in high energy hadronic collisions is factorized into a single
 collision between two partons multiplied by a probability to find
these two partons of momentum fractions $x_1$ and $x_2$, respectively,
from two incoming hadrons.  The probability is then factorized into a
product of two parton distributions $\phi(x_1)$ and $\phi(x_2)$, 
which are probabilities to find these two partons from the respective 
hadrons \cite{Factorization}.
Because of extremely large number of soft gluons in heavy 
ion beams, it is natural to go beyond
the factorized single-scattering formalism to include any possible
multiple scattering, and long range correlations between soft gluons
from two incoming ions.

McLerran and Venugopalan (MV) developed a new formalism for 
calculation of the soft gluon distribution for very large 
nuclei \cite{Raju1,Raju2}. In this approach, the valence quarks in the 
nuclei are treated as the classical source of the color charges.
They argued that the valence quark recoil can be ignored in the limit
when the gluons emitted are soft.  
In addition, because of the Lorentz contraction, the color charge 
of the valence quarks is treated approximately as an infinitely 
thin sheet of color charge along the light cone. 
With these assumptions, the gluon distribution function for very large 
nuclei may be obtained by solving the classical Yang-Mills Equation
\cite{Raju2,Raju3}. Using the classical glue field generated by a
single nucleus obtained in the MV formalism as the basic
input, Kovner, McLerran, and Weigert (KMW)
computed the soft gluon production in a collision of two 
ultra-relativistic heavy nuclei by solving the classical  
Yang-Mills equations with the iteration to the second order \cite{KMH}. 
The two nuclei are treated as the infinitely thin sheets of the 
classical color charges  moving at the speed of light in the positive 
and the negative $z$ directions, respectively.   Following this 
approach, the distribution of soft gluons at the rapidity $y$ and the
transverse momentum $p_T$ in nuclear collisions can be expressed as
\cite{KMH,MGLM}
\begin{equation}
\frac{d \sigma}{dy d^2 p_T}=
             \frac{2g^6}{(2\pi)^4}
             \left(\frac{N_q}{2N_c}\right)^2 
              N_c (N_c^2-1)\, \frac{1}{p_T^4}\, 
    \ell n \left(\frac{p_T^2}{\Lambda^2_{cutoff}}\right)\ .
\label{ae0}
\end{equation}
where $g$ is the strong coupling constant, $N_c=3$ is the number
of the color, and $\Lambda_{cutoff}$ is a cutoff mass scale \cite{KMH}.  
In Eq.~(\ref{ae0}), $N_q/(2N_c)=S_T\mu^2$, where $\mu^2$ is the
averaged color charge squared per unit area of the valence quark, and
$S_T$ is the transverse area of the nuclei. $N_q$ is the total number of the 
quarks in the color charge source. Eq.~(\ref{ae0})) is potentially 
very useful in estimating the production of mini-jet rates, and the formation 
of the possible quark-gluon plasma at RHIC \cite{GW}.

In the following, I show that this result can be understood in terms of 
QCD perturbation theory. I will also  explore under
what kind of approximation this result matches the conventional 
perturbative calculation \cite{Guo}.  

\section[Partonic process $qq\rightarrow qqg$]
{Partonic process $qq\rightarrow qqg$}

\begin{figure}
\epsfig{figure=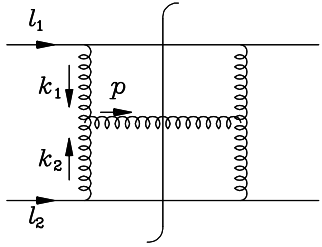,width=2.0in}
\caption{Square of the leading Feynman diagram to the process: 
$qq\rightarrow qqg$.}
\label{fig1}
\end{figure}

KMW's derivation for Eq.~(\ref{ae0}) is based on the following physical 
picture: in ultra-relativistic heavy ion collisions, gluons are 
produced by the fields of two strongly Lorentz contracted color 
charge sources, which are effectively equal to the valence quarks of
two incoming ions.  In order to understand KMW's result in 
the language of perturbative QCD, let us consider a specific partonic 
process: $qq\rightarrow qqg$, as sketched in Fig.~\ref{fig1}.
If we assume that the incoming quarks $qq$ are the valence quarks in
the initial color charge sources, the partonic subprocess in
Fig.~\ref{fig1} mimics the physical picture adopted in KMW's
derivation. In this section, I show how to extract the leading 
contribution to the soft gluon production from this diagram. 
However, as a Feynman diagram in QCD perturbation theory,
the single diagram shown in Fig.~\ref{fig1} is not gauge invariant.
In next section, I show that with a proper choice of gauge and certain 
approximation, other diagrams are suppressed in soft gluon limit.  

In general, the invariant cross section of the gluon production,  
shown in Fig.~\ref{fig1}, can be written as
\begin{equation}  
d\sigma_{qq\rightarrow g} / dy d^2 p_T 
= \frac{1}{2s}\, |\overline{M}|^2\, dps \ ,
\label{e2}
\end{equation}
where $s=(l_1+l_2)^2$, and $|\overline{M}|^2$ is matrix
element square with the initial-spin averaged and the final-spin summed.
$l_1$ and $l_2$ as the momenta of the two incoming quarks respectively.
In Eq.~(\ref{e2}), the phase space 
\begin{eqnarray}
dps & \propto & 
\frac{d^4 k_1}{(2\pi)^4} (2\pi) \delta ((l_1-k_1)^2)
    \times \frac{d^4 k_2}{(2\pi)^4} (2\pi) \delta ((l_2-k_2)^2)
 \nonumber \\
& &  \times (2\pi)^4 \delta^4 (k_1+k_2-p) \ ,
\label{e3}
\end{eqnarray}
where $k_1$ and $k_2$ are the momenta for the two  
gluons emitted from the initial quarks.  Because of the 
gluon propagators, as shown in Fig.~\ref{fig1}, the matrix element
square $|\overline{M}|^2$ has the following pole structure:  
\begin{equation}
poles=\frac{1}{k_1^2+i\epsilon}\, \frac{1}{k_1^2-i\epsilon}\,
      \frac{1}{k_2^2+i\epsilon}\, \frac{1}{k_2^2-i\epsilon}\, .
\label{e4}
\end{equation}
When integrating over the phase space, we see that the leading
contribution comes from the terms with $k_1^2 \rightarrow 0$ or
$k_2^2 \rightarrow 0$ limit. Note that  $k_1^2$  and $k_2^2$ can 
not be zero at the same time,  because
$k_1$ and $k_2$ come from different directions, and we have the 
on-shell condition of $p^2=(k_1+k_2)^2=0$.  
Therefore, to calculate the leading contribution, we can
first calculate the diagram in $k_1^2 \rightarrow 0$ limit.  The total
leading contribution is just twice of it, because the diagram is
symmetric for $k_1$ and $k_2$. 
\begin{figure}
\begin{minipage}[t]{2.5in}
\epsfig{figure=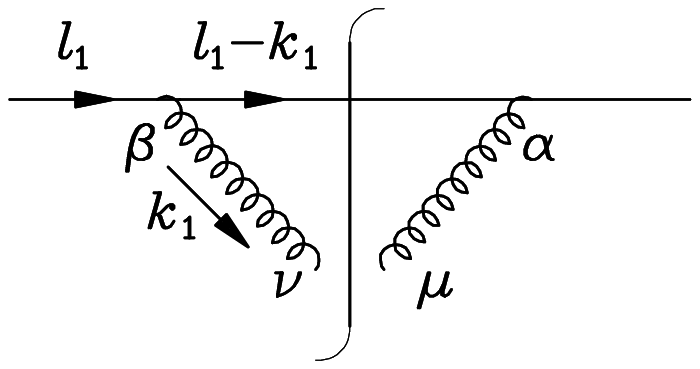,width=2.0in}
\caption{Feynman diagram for quark splitting: 
$P_{l_1 \rightarrow k_1}(x, k_{1T}<p_T)$. }
\label{fig2}
\end{minipage}
\hfill
\begin{minipage}[t]{2.5in}
\epsfig{figure=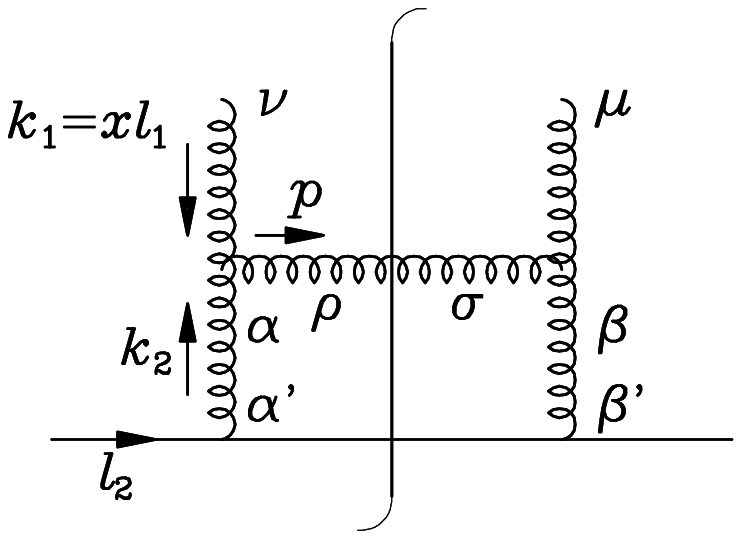,width=2.0in}
\caption{Leading Feynman diagram contributing to the hard scattering 
part $H(xl_1,l_2,p)$. }
\label{fig3}
\end{minipage}
\end{figure}

To derive the leading contribution at $k_1^2 \rightarrow 0$ limit, 
we perform the collinear approximation
$k_1\approx xl_1+O(k_{1T})$, with $k_{1T} \sim \Lambda_{cutoff} <<
p_T$, where $\Lambda_{cutoff}$ is a collinear cutoff scale 
\cite{GS:TASI}.  This approximation means that the leading 
contribution is from the phase space where almost all 
transverse momentum of the final-state gluon comes from the gluon of
$k_2$, and $k_1$ is almost collinear to $l_1$.  After such collinear 
approximation, the cross section in Eq.~(\ref{e2}) can be 
approximately written in a factorized form \cite{Factorization}: 
\begin{eqnarray}
\frac{d\sigma_{qq\rightarrow g}}{dydp_T^2} 
&\approx & 2\left(\frac{1}{2(2\pi)^3}\, \frac{1}{2s} \right)
            \int \frac{dx}{x} \, 
             P_{l_1 \rightarrow k_1} (x, k_{1T}<p_T) \, 
             H(xl_1,l_2,p)
\nonumber \\
&\ & + O(\frac{\Lambda_{cutoff}^2}{p_T^2})  \, ,
\label{e5}
\end{eqnarray}
where the overall factor of 2 is due to the fact that  the leading 
contribution come from two regions corresponding to 
$k_1^2\rightarrow 0$ and $k_2^2\rightarrow 0$, respectively. 
In Eq.~(\ref{e5}), $P_{l_1 \rightarrow k_1}(x, k_{1T}<p_T)$ represents
the probability of finding an almost collinear gluon with the momentum
fraction $x$ from an incoming quark of momentum $l_1$, and is 
represented by Fig.~\ref{fig2}.  $H(xl_1,l_2,p)$ in Eq.~(\ref{e5})
is effectively the hard scattering between the gluon of $k_1=xl_1$ 
and the incoming quark of $l_2$, and it is represented by the diagram 
shown in Fig.~\ref{fig3}.  

Under the collinear expansion $k_1\approx xl_1$, the gluon line which 
connects the partonic parts $P_{l_1 \rightarrow k_1}$ and 
$H(xl_1,l_2,p)$ is effectively on the mass-shell, and therefore, the 
partonic parts, $P$ and $H$ in Eq.~(\ref{e5}), are separately gauge 
invariant. The quark splitting function $
P_{l_1 \rightarrow k_1}(x,k_{1T}<p_T)$ 
can be calculated in $n\cdot A=0$ gauge. We have \cite{Guo} 
\begin{eqnarray}
&\ & P_{l_1 \rightarrow k_1}(x,k_{1T}<p_T) 
\nonumber \\
& =& \frac{N_c^2-1}{2N_c} \,  
    \left( \frac{g^2}{8\pi^2} \frac{1+(1-x)^2}{x} \right)
      \ell n \left(\frac{p_T^2}{\Lambda^2_{cutoff}}\right)\, .
\label{e13}
\end{eqnarray}
In $\bar{n}\cdot A=0$ gauge,  we have the hard scattering 
function \cite{Guo}
\begin{eqnarray}
H(xl_1,l_2,p) &=& (2\pi) 4g^4 \left(\frac{1}{2}\right)
        \left( \frac{p_+}{xl_+} \right) ^2 \frac{1}{p_T^4}
       \frac{1}{s-2l_+p_-} 
       \delta (x-\frac{2p_+l_-}{s-2l_+p_-} ) \nonumber \\
&\ & \times
       \left[ (xs-2xl_+p_-)^2+(2xl_+p_-)(2p_+p_-) \right]\, .
\label{e21}
\end{eqnarray}
For soft gluon production, we define the soft gluon limit as
\begin{equation}
\frac{p_{-}}{l_{-}} \ll 1 \,  
\quad \mbox{and} \quad
\frac{p_{+}}{l_{+}} \ll 1 \, .
\label{e23}
\end{equation}
Combining Eq.~(\ref{e13}), Eq.~(\ref{e21}) and Eq.~(\ref{e5}), 
and taking the soft gluon limit, we have \cite{Guo}
\begin{equation}
\frac{d\sigma_{qq\rightarrow g}}{dy d^2 p_T}
  = \frac{2g^6}{(2\pi)^4}
    \left(\frac{1}{2N_c}\right)^2 
    N_c (N_c^2-1)\left( \frac{1}{p_T^4} \right) 
    \ell n \left(\frac{p_T^2}{\Lambda ^2 _{cutoff}}\right)\, .
\label{e26}
\end{equation}
Eq.~(\ref{e26}) reproduced Eq.~(\ref{ae0}), which is obtained by 
solving classical Yang-Mills equations, at $N_q=1$. If we consider 
the total number of the quarks in the charge sources of both sides, 
we need to multiply $N_q^2 $ to Eq.~(\ref{e26}). 

\section{Gauge Invariance}

\begin{figure}
\epsfig{figure=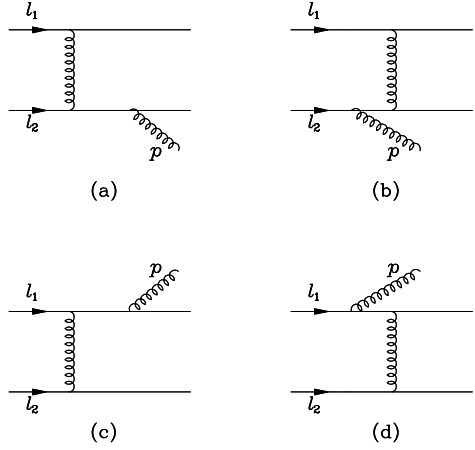,width=2.0in}
\caption{The rest of Feynman diagrams contributing to the process: 
$qq\rightarrow qqg$, in addition to the diagram in Fig.~1.}
\label{fig4}
\end{figure}

\begin{figure}
\epsfig{figure=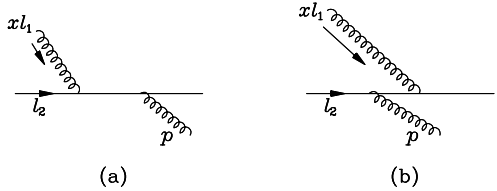,width=2.0in}
\caption{Feynman diagrams contributing to $H_a(xl_1,l_2,p)$ (a),
and $H_b(xl_1,l_2,p)$ (b). }
\label{fig5}
\end{figure}

For the gluon production in the process $qq\rightarrow qqg$, 
the single diagram shown in Fig.~\ref{fig1} is not gauge invariant. 
In general, we also need to consider the radiation diagrams shown in
Fig.~\ref{fig4}.  
Similar to the diagram in Fig.~\ref{fig1}, 
the contribution of Fig.~\ref{fig4}a and Fig.~\ref{fig4}b 
can also be written in the same factorized form:
\begin{eqnarray}
E \frac{d\sigma_{qq\rightarrow g}^{\rm rad}}{d^3 p} 
&\approx  & \frac{1}{2(2\pi)^3}\, \frac{1}{2s} \,
            \int \frac{dx}{x}\, 
             P_{l_1 \rightarrow k_1} (x, k_{1T}<p_T) \, 
             H_{i}(xl_1,l_2,p)
\nonumber \\
&\ &     +  O(\frac{\Lambda_{cutoff}^2}{p_T^2})  \, ,
\label{ae1}
\end{eqnarray}
with $i=a, b$. $P_{l_1 \rightarrow k_1} (x, k_{1T}<p_T)$ is the 
quark splitting function given by Eq.~(\ref{e13}). 
$H_{a}(xl_1,l_2,p)$ and 
$H_{b}(xl_1,l_2,p)$ are the hard scattering parts from the diagrams in 
Fig.~\ref{fig4}a and Fig.~\ref{fig4}b, and they are represented by
Fig.~\ref{fig5}a and Fig.~\ref{fig5}b, respectively. 
With the contribution from diagrams in Fig.~\ref{fig4}, 
Eq.~(\ref{e5}) changes to
\begin{eqnarray}
E \frac{d\sigma_{qq\rightarrow g}}{d^3 p} 
   & \approx & 2\left(\frac{1}{2(2\pi)^3}\, \frac{1}{2s} \right)
            \int \frac{dx}{x}\, 
            P_{l_1 \rightarrow k_1} (x, k_{1T}<p_T) 
\nonumber \\
&\ & \quad\quad\quad \times 
          \left[ H(xl_1,l_2,p)
                +H_{a}(xl_1,l_2,p)
                +H_{b}(xl_1,l_2,p) \right. 
\nonumber \\
&\ &  \quad\quad\quad
         \left.   +\mbox{interference terms} \right] 
         +  O(\frac{\Lambda_{cutoff}^2}{p_T^2})  \, ,
\label{ae3}
\end{eqnarray}
with the approximation $k_1 =xl_1 +O(k_{1T})$.
Feynman diagrams  shown in Fig.~\ref{fig3} and Fig.~\ref{fig5} form 
a gauge invariant subset for calculating the hard scattering parts,
$H(xl_1,l_2,p)$'s in Eq.~(\ref{ae3}). With our choice of gauge, 
$H_i/H \sim p_{-}/l_{-} << 1$ in the soft gluon limit. Therefore, 
the contribution from diagrams in Fig.~\ref{fig4}a and Fig.~\ref{fig4}b
can be neglected in comparison with the contribution from the 
diagram in Fig.~\ref{fig1} at $k_1^2 \sim 0$. 

Similarly, the contributions from diagrams shown in  Fig.~\ref{fig4}c and  
Fig.~\ref{fig4}d can be neglected in the soft gluon limit, when compared 
with the contribution from the diagram in Fig.~\ref{fig1} at 
$k_2^2 \sim 0$. Therefore, with the approximation of 
$k_1^2\sim 0$ (or $k_2^2\sim 0$) 
and the soft gluon limit, and a proper choice of the gauge, 
the contributions {\it extracted} from the diagram in
Fig.~\ref{fig1} to the leading soft gluon production in Eq.~(\ref{ae0})
is gauge invariant.

\section{Discussion}

When we consider the collision between two nuclei, we can treat the two 
incoming quarks in Fig.~\ref{fig1} as coming from two nuclei respectively. 
In this picture, the number of the valence quark $N_q$ is replaced by the
quark distribution in the nuclei.
In terms of the parton model, 
the cross section between the two nuclei $A$ and $B$ can be 
expressed in the following form:
\begin{equation}
\frac{d\sigma_{AB\rightarrow g}}{dy dp_T^2} = 
         \int dz_1\, dz_2\, f_{q/A}(z_1)\, f_{q/B}(z_2)\,
         E \frac{d\sigma_{qq\rightarrow g}}{d^3 p}\, .
\label{ae4}
\end{equation}
Here $z_1$ and $z_2$ are the momentum fractions of the quarks, and 
$f_{q/A}(z_1)$ and  $f_{q/B}(z_2)$ are the quark distributions (or quark 
number densities) of the two nuclei.  If we denote
$p_A$ and $p_B$ as the momenta for the two nuclei respectively, then
$z_1=l_1/p_A$  and $z_2=l_2/p_B$. Substituting Eq.~(\ref{e5}) into
Eq.~(\ref{ae4}), we have
\begin{eqnarray}
 \frac{d\sigma_{AB\rightarrow g}}{dy dp_T^2} 
&\approx &  
    \frac{1}{2(2\pi)^3}\, \frac{1}{2S} \, 
    \int \frac{dz_1}{z_1} \, \frac{dz_2}{z_2} 
\nonumber \\
&\ & \times 
\Bigg[
    \int \frac{dx_1}{x_1}\, f_{q/A}(z_1) f_{q/B}(z_2)
 \nonumber \\
&\ & {\hskip 0.2in} \times
   P_{l_1 \rightarrow k_1} (x_1, k_{1T}<p_T) \,   H(x_1l_1,l_2,p)
\nonumber \\
&\ & +  
    \int \frac{dx_2}{x_2}\, f_{q/A}(z_1) f_{q/B}(z_2) 
\nonumber \\
&\ & {\hskip 0.1in} \times
    P_{l_2 \rightarrow k_2} (x_2, k_{2T}<p_T) \, H(l_1,x_2l_2,p) 
    \Bigg]
\label{ae5}
\end{eqnarray}
where the overall factor 2 in Eq.~(\ref{e5}) is now represented by 
the two terms, and $S=(p_A+p_B)^2\approx 2p_A \cdot p_B$.  
In Eq.~(\ref{ae5}), $x_i=k_i/l_i$ with $i=1,2$, 
and $l_1=z_1p_A$ and $l_2=z_2p_B$. If we denote the 
momentum fraction of gluon $k_1$ with respect to $p_A$ 
as $z_1'=k_1/p_A$, and $k_2$ with respect to $p_B$ 
as $z_2'=k_2/p_B$, we can rewrite Eq.~(\ref{ae5}) in terms of 
$z_1'$ and $z_2'$:
\begin{eqnarray}
\frac{d\sigma_{AB\rightarrow g}}{dy dp_T^2} 
& \approx & 
    \frac{1}{2(2\pi)^3}\, \frac{1}{2S} 
\nonumber \\
&\ & \times  \Bigg\{ \int \frac{dz_1'}{z_1'} \, \frac{dz_2}{z_2}
     \left[ \int \frac{dz_1}{z_1}\, f_{q/A}(z_1)\,
            P_{l_1 \rightarrow k_1} (z'/z_1, k_{1T}<p_T) \right]
\nonumber \\
&\ & {\hskip 0.2in} \times
     f_{q/B}(z)\, H(z_1'p_A,z_2p_B,p) 
\nonumber \\
&\ & +\int \frac{dz_2'}{z_2'} \, \frac{dz_1}{z_1}
     \left[ \int \frac{dz_2}{z_2}\, f_{q/B}(z_2)\,
            P_{l_1 \rightarrow k_1} (z_2'/z_2, k_{2T}<p_T) \right]
\nonumber \\
&\ & {\hskip 0.2in} \times 
     f_{q/A}(z_1)\, H(z_1p_A,z_2'p_B,p) \Bigg\}
\label{ae6p}
\end{eqnarray} 
According to the QCD factorization theorem \cite{Factorization}, 
we see that the part inside the square
brackets is actually the gluon distribution from nuclei $A$ 
(or $B$) at 
the factorization scale $\mu_F^2=p_T^2$, with only the quark 
splitting function \cite{Field}, 
\begin{eqnarray}
f_{g/A}(z_1',\mu_F^2=p_T^2) &=& 
        \int \frac{dz_1}{z_1}\, f_{q/A}(z_1) 
        P_{l_1 \rightarrow k_1} (z_1'/z_1, k_{1T}<p_T) 
\nonumber \\
&\ & +   \mbox{term from gluon splitting}.
\label{ae7}
\end{eqnarray}
Using Eq.~(\ref{ae7}), we can then reexpress Eq.~(\ref{ae6p}) as:
\begin{eqnarray}
\frac{d\sigma_{AB\rightarrow g}}{dy dp_T^2} 
 & \approx &
        \frac{1}{2(2\pi)^3}\, \frac{1}{2S} \,
\nonumber \\ 
&\ &  \times 
        \int \frac{dz'}{z'} \, \frac{dz}{z}  
         \left[ f_{g/A}(z',\mu_F^2=p_T^2) f_{q/B}(z) 
                H(z'p_A,zp_B,p) \right.
\nonumber \\
&\ & {\hskip 0.1in} \left. 
              + f_{q/A}(z) f_{g/B}(z',\mu_F^2=p_T^2) 
                H(zp_A,z'p_B,p) \right]
\, ,
\label{ae8}
\end{eqnarray} 
which is the factorized formula for  two-to-two subprocesses in 
the conventional perturbative QCD for the nucleus-nucleus collisions.
In KMW formalism, only the valence quark color charge was used 
as the source of the classical charge of colors.  As a result,  
the gluon splitting term in Eq.~(\ref{ae7}) is neglected for the 
distribution $f_{g/A}$.

From the above comparison, we conclude that 
by solving the classical Yang-Mills Equation to the second order in 
iteration, KMW's result is consistent with conventional 
perturbative QCD at the leading logarithmic approximation. 
The logarithmic dependence shown in KMW's result basically 
describes the logarithmic DGLAP evolution of the quark 
distributions \cite{GLAP}. However, the iteration 
in this approach is different from the expansion series in 
conventional perturbative QCD. The McLerran-Venugopalan 
formalism was later further developed to include 
the harder gluons into the charge density $\mu^2$
and treat the charge source as an extended distribution which depends 
on the rapidity \cite{JKMW}. It will be potentially very useful 
if this new approach, after including higher orders of iteration, 
can include the parton recombination \cite{recom} 
and other non-perturbative effects which are not 
apparent in the normal perturbative calculation.

\end{document}